\documentclass[11pt]{article}
\usepackage{graphicx}
\usepackage{setspace}
\usepackage{array}

\usepackage{amsmath}
\usepackage{epstopdf}
\usepackage{authblk}
\usepackage{cite}
\usepackage{fullpage}
\usepackage{float}
\title{\bf Numerical and analytical studies on the complete synchronization of the variant of Murali-Lakshmanan-Chua circuits}
\author[1]{M.Daniel Sweetlin}
\affil{Department of Physics, St.John's College, Palayamkottai-627 002, India}
\author[2]{G.Sivaganesh \footnote {Corresponding author : sivaganesh.nld@gmail.com}}
\affil{Department of Physics, Alagappa Chettiar College of Engineering $\&$ Technology, Karaikudi-630 004, India}
\date{\today}
\begin{document}
\maketitle
\begin{abstract}
In this paper we present numerical and analytical studies on the complete synchronization phenomena exhibited by unidirectionally coupled two variant of {\emph{Murali-Lakshmanan-Chua}} circuits. The transition of the coupled system from an unsynchronized state to a state of complete synchronization under the influence of the coupling parameter is observed through phase portraits obtained numerically and analytically.
\end{abstract}
\
%\doublespace
\section{Introduction}

The phenomenon of chaos synchronization in coupled chaotic systems has been studied extensively for the past two decades with a motivation to understand the coherent dynamical behaviour of coupled systems. Chaos synchronization in nonlinear electronic circuits has been observed in a variety of circuits \cite{Chua1992705,Chua199393,Murali19931624,Murali19944882,Murali1995563,Lakshmanan1996,Murali1997415} and finds potential applications in secure communication. Several synchronization phenomena such as complete, phase and lag synchronization have been identified in identical and non-identical chaotic systems and studied both experimentally and numerically. A detailed study on characterizing the above said synchronization phenomena was done numerically \cite{Boccaletti20021}. Complete synchronization being the strongest among the other types of synchronization occurs when the states of the coupled systems converge, irrespective of the mismatch in initial conditions. Complete synchronization phenomena exhibited by coupled second order dissipative electronic circuits was studied experimentally, numerically \cite{Lakshmanan1996, Murali1995563} and analytically \cite{Sivaganesh2015010503}. The synchronization of coupled Chua circuits and coupled $Murali-Lakshmanan-Chua(MLC)$ circuits  was studied numerically \cite{Murali1997415}. The $MLC$ circuit being the simplest diisipative non-autonomous chaotic circuit was introduced by Murali \cite{Murali1994462}. The period doubling dynamics of the circuit leading to chaotic motion was studied both experimentally and numerically \cite{Murali19941511,Lakshmanan199533}. An explicit analytical solution to the normalized circuit equations of the $MLC$ circuit were presented \cite{Lakshmanan1996,Lakshmanan199533,Lakshmanan2003}. The $MLC$ circuit has the picewise linear nonlinear element, the Chua's diode, which is linear within the three regions. Since the circuit equations describing the $MLC$ circuit in each of the piecewise linear region is a second order differential equation, the equation is solved for each of the linear regions and matched across the boundaries. Similar solutions were given to some simple  second order nonlinear chaotic circuits with piecewise nonlinear element as the active circuit component \cite{Thamilmaran2001783,Thamilmaran2005637,Manimehan20092347, Arulgnanam20092246} and for some circuits exhibiting Strange Nonchaotic Attractors {\emph{(SNA)}} in their dynamics \cite{Sivaganesh20141760}. The complete synchronization phenomenon exhibited by undirectionally coupled two $MLC$ circuits was studied analytically through phase portraits generated from the analytical solution \cite{Sivaganesh2015010503}. \\
The variant of the $MLC$ circuit $(MLCV)$ is a simple forced parallel $LCR$  circuit with the Chua's diode as the nonlinear element, connected parallel to the capacitor, was suggested by Thamilmaran \cite{Thamilmaran20001175}. The quasiperiodic route and the reverse period doubling route to chaos exhibited by $MLCV$ circuit has been extensively studied \cite{Thamilmaran20001175,Thamilmaran2001783}. Further an explicit analytical solution to the state variables of the normalized circuit equations are obtained. The solutions thus obtained were used to explain the dynamics of the circuit through phase portraits of the state variables. The complete synchronization phenomena exhibited by $MLCV$ circuit was studied experimentally \cite{ThamilmaranPhd06}. However the dynamics of the coupled $MLCV$ circuit has not been yet studied numerically and analytically. In this paper, we study the complete sychronization phenomenon exhibited by unidirectionally coupled two variant of {\emph{Murali-Lakshmanan-Chua}} circuits numerically and provide an explicit analytical solution for the state variables of the normalized circuit equations characterizing the coupled system. 

\subsection*{Circuit Equations}
The variant of the {\emph{Murali-Lakshmanan-Chua}} circuit is a simple forced parallel $LCR$ circuit with the Chua's diode as the only nonlinear element, connected parallel to the capacitor $C$. The normalized state equations of the coupled circuit is given as,
\vspace{0.1cm}
\begin{subequations}
\begin{eqnarray}
\dot x & = & f_1 sin(\omega_1 t) - x - y - g(x), \\
\dot y & = & \beta x  
\end{eqnarray}
\end{subequations}
The piecewise linear function $g(x)$ representing the Chua's diode is given by,
\begin{equation}
g(x) =
\begin{cases}
bx+(a-b) & \text{if $x\ge 1$}\\
ax & \text{if $|x|\le 1$}\\
bx-(a-b) & \text{if $x\le -1$}
\end{cases}
\end{equation}
where, $ \beta = (C/LG^2)$, $ a  = G_a/G$,  $ b = G_b/G$, $f_1 = (F_1 \beta/B_p)$, $\omega_1 = (\Omega_1 C/G)$ and  $ G = 1/R$. The circuit parameters takes the values $C=10.15 nF$, $L=445 mH$, $R=1475 \Omega$ and the parameters of the Chua's diode are chosen as $G_a = -0.76 mS$, $G_b = -0.41 mS$ and $B_p = 1 V$. The frequency of the external periodic force  $\nu_1$ is fixed as $\nu_1 = \Omega/2\pi = 1.116 kHz$. 
\begin{figure}[t]
\begin{center}
\includegraphics[scale=0.6]{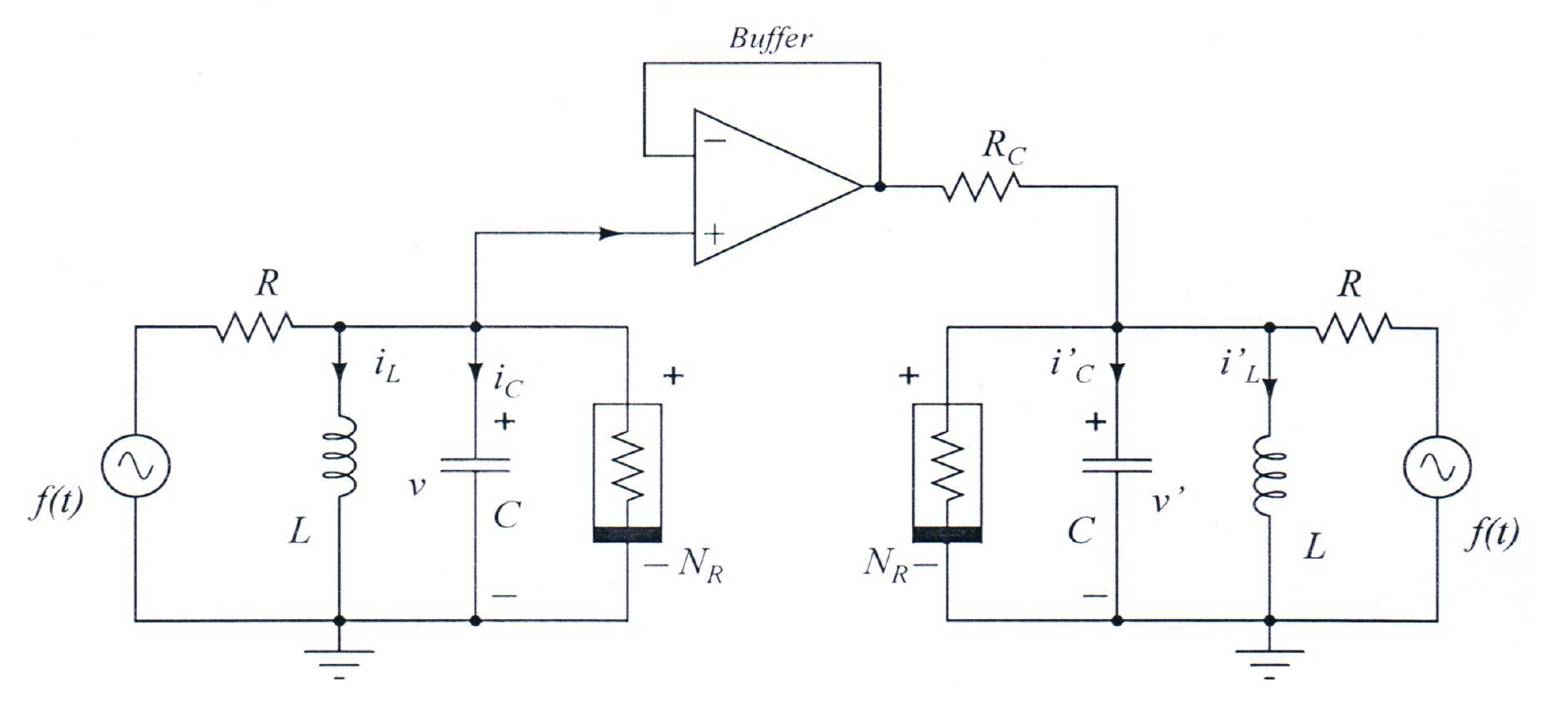}
\caption{ Schematic circuit realization of two identical MLCV circuits with unidirectional coupling}
\label{fig:1}
\end{center}
\end{figure}
The $MLCV$ circuit explained above acting as the drive system is unidirectionally coupled to another $MLCV$ circuit which acts as the response system. The drive and the response systems operate with different initial conditions, different values of amplitude of external periodic forcing, are coupled by a linear resistor and a buffer. The buffer acts as a signal driving element which isolates the drive system variables being affected by the response system.The schematic diagram of the coupled $MLC$ circuits is as shown in Fig.1. The normalized state equations of the response system is given as,
\begin{subequations}
\begin{eqnarray}
\dot {x^{'}} & = & f_2 sin(\omega_2 t) - x^{'} - y^{'} - g(x^{'})+\epsilon(x-x^{'}),\\
\dot {y^{'}} & = & \beta x^{'} 
\end{eqnarray}
\end{subequations}
and the piecewise linear function $g(x^{'})$ is given by,
\begin{equation}
g(x^{'}) =
\begin{cases}
bx^{'}+(a-b) & \text{if $x^{'}\ge 1$}\\
ax^{'} & \text{if $|x^{'}|\le 1$}\\
bx^{'}-(a-b) & \text{if $x^{'}\le -1$}
\end{cases}       
\end{equation}
where, $f_2 = (F_2 \beta/B_p)$, $\omega_2 = (\Omega_2 C/G)$  and $\epsilon = (R/R_c)$ is the coupling parameter. The circuit parameters of drive and response systems are fixed as $C=10.15 nF$, $L=445 mH$, $R=1475 \Omega$ and the parameters of the Chua's diode are chosen as $G_a = -0.76 mS$, $G_b = -0.41 mS$ and $B_p = 1 V$.

\section{Numerical Analysis}

In this section, we provide the numerical simulation results of the unsynchronized and synchronized states of the coupled system. Further, we present the eigenvalues of the difference system, obtained from the normalized state equations of the system, as a function of the coupling paramter. \\
From the normalized state equations of the response system given by Eqs.(3) we could infer that the two circuits are independent of each other when the coupling resistor $R_c$  takes very larger values such that the coupling parameter $\epsilon = R/R_c = 0$. Hence, the coupling resistor blocks the response being controlled by the drive and makes it independent of the drive system. The frequency of the external periodic force are fixed as $\omega_{1,2} = 0.105$ while their correponding amplitudes are fixed to have different values as $f_1 = 0.38$ and $f_2 = 0.381$. In addition to that, the drive and response systems are operated with different set of initial conditions as $x_0=-0.5, y_0=0.1$ and $x^{'}_0 = 0.5, y^{'}_0=0.11$. It is to be expeceted that the chaotic attractors of the drive and response systems must be unsynchronized because of the mismatch in initail conditions and external forcing.  Fig.2(a), (b) shows the unsynchronized state of the drive and the response systems in the $x-x^{'}$ phase plane and the trajectory plot in $x-x^{'}$ plane respectively for $\epsilon=0$. However, for small values of $R_c$, $ \epsilon = R/R_c > 0$, the coupling resistor allows signal flow from drive to response and the two systems are strongly coupled. Owing to unidirectional coupling, the dynamics of the drive is unaltered while that of the response system varies with the coupling parameter $\epsilon$ and hence it is controlled by the drive. Fig.2(c),(d) shows the synchronized state of the two systems in the  $x-x^{'}$ phase plane and the trajectory plot in $x-x^{'}$ plane respectively for $\epsilon=1$. It is found that the drive and the response systems which are unsynchronized for low values of coupling paramter $\epsilon$ are completely synchronized for higher values.\\
\begin{figure}[htb]
\begin{center}
\includegraphics[scale=0.66]{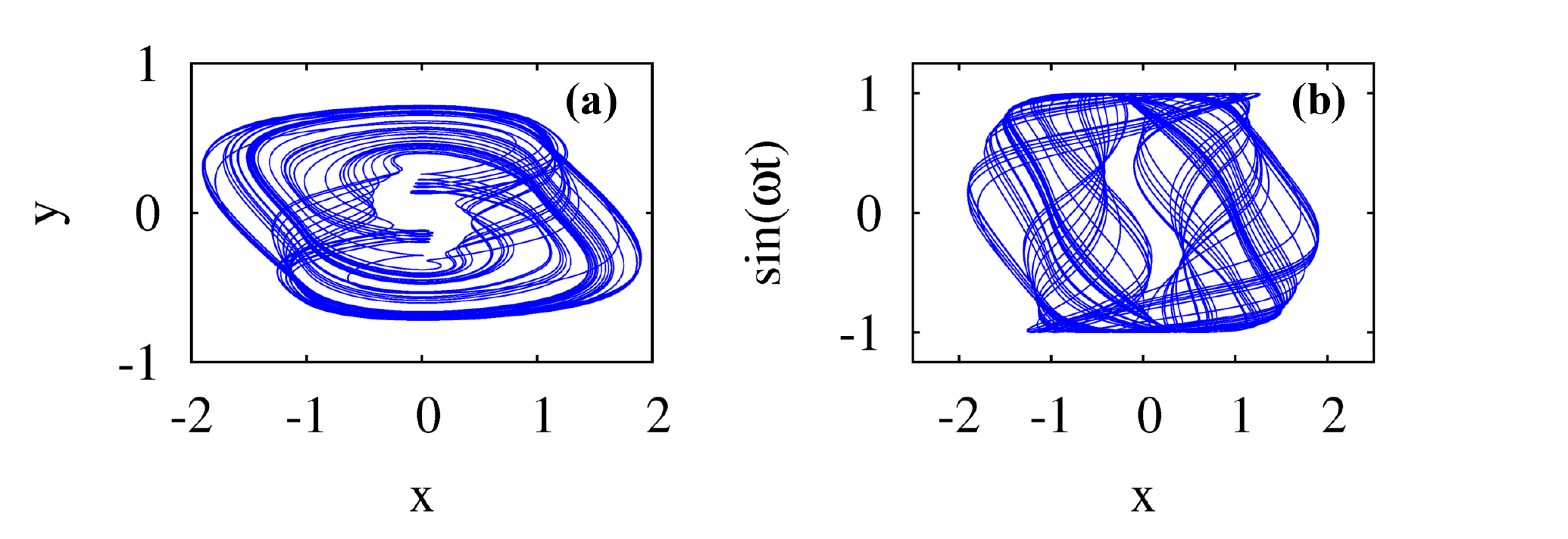}
\caption{Numerically obtained chaotic attractors in (a) $x-y$ (b) $x-sin(\omega_1 t)$ phase planes.}
\label{fig:2a}
\end{center}
\end{figure}
\begin{figure}[htb]
\begin{center}
\includegraphics[scale=0.6]{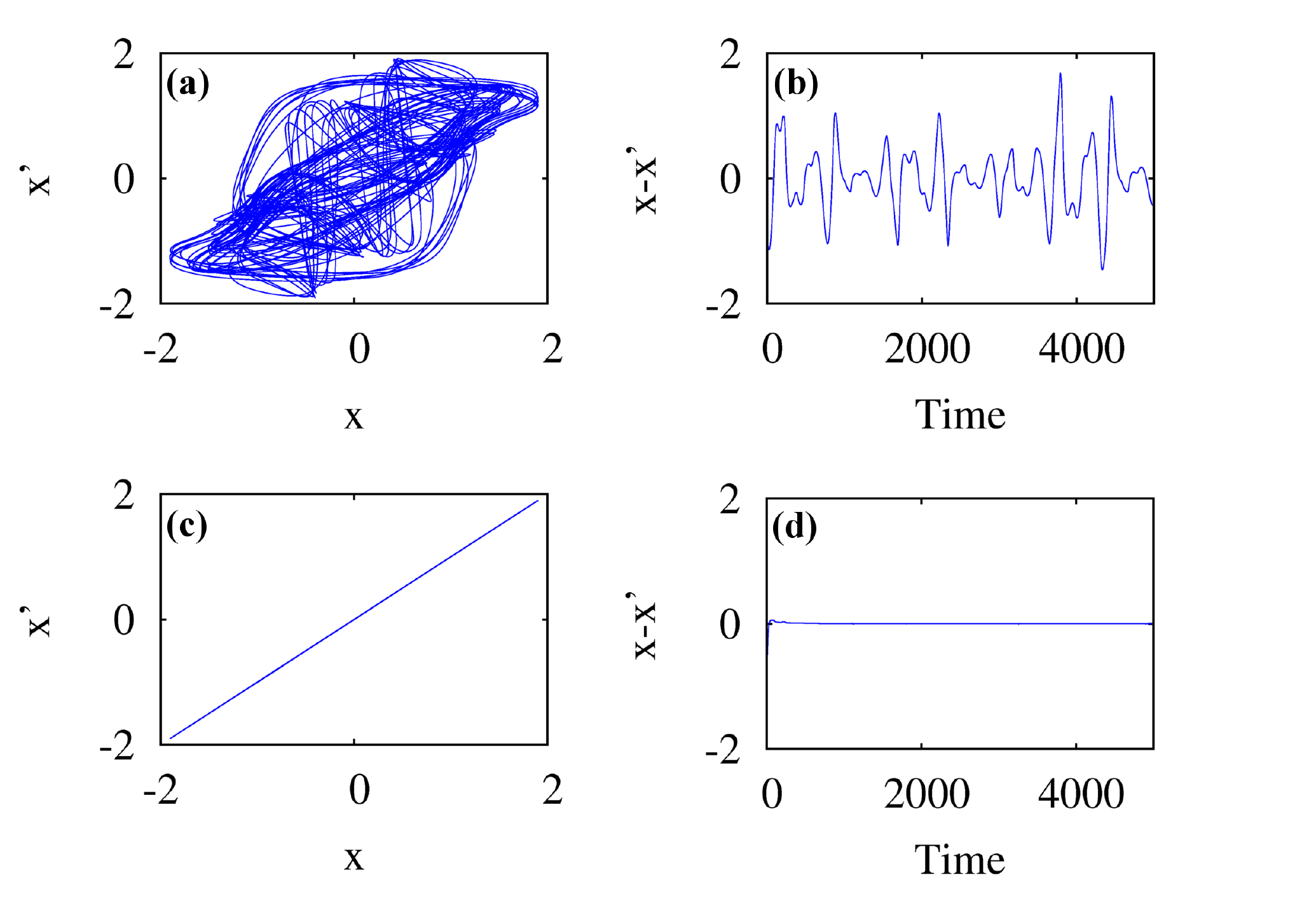}
\caption{(a) Unsynchronized motion in $(x-x^{'})$ plane for $\epsilon = 0.0$, $f_1 = 0.38$ and $f_2 = 0.381$, [$x_0$=-0.5, $y_0$=0.1] and [$x^{'}_0$ = 0.5, $y^{'}_0$=0.11]. (b) Trajectory plot in $(x-x^{'})$ plane. (c) Synchronized motion in $(x-x^{'})$ plane for $\epsilon = 1$, $f_1 = 0.38$ and $f_2 = 0.381$, [$x_0$=-0.5, $y_0$=0.1] and [$x^{'}_0$ = 0.5, $y^{'}_0$=0.11]. (d) Trajectory plot in $(x-x^{'})$ plane.}
\label{fig:2a}
\end{center}
\end{figure}
\subsection*{Stability Analysis}

From the numerical results obtained, we observe that the dynamics of the response is influenced by the drive. Hence, the eigenvalues of the response system must depend upon the coupling parameter. Since the drive and response systems are piecewise linear, each picewise linear regions of the two sytems could be coupled together to get a new set of equations. In this section, the dependence of the eigenvalues and their stability on the coupling paramete for each of the piecewise linear regions is presented. The difference system obtained from Eqs.(1) and (3) are
\begin{subequations}
\begin{eqnarray}
\dot {x^{*}} & = & f_1 sin(\omega_1 t) - f_2 sin(\omega_2 t) - x^{*} - y^{*}- (g(x) -g(x^{'}))-\epsilon x^{*},\\
\dot {y^{*}} & = & \beta x^{*} 
\end{eqnarray}
\end{subequations}
where $x^{*}$=$(x-x^{'})$, $y^{*}$=$(y-y^{'})$ and $g(x) -g(x^{'})=g(x^{*})$ takes the values $a{x^{*}}$ or $b{x^{*}}$ depending upon the corresponding region of operation of the drive and response systems. Since the circuit equations are piecewise linear, the equations representing the identical regions of the two circuits are coupled to get a new set of equation for the corresponding regions. The state variables of this new set of equations, $x^{*}(t),~y^{*}(t)$ are given as,
\begin{subequations}
\begin{eqnarray}
x^{*} = x -  x^{'} \\ 
y^{*} = y -  y^{'},
\end{eqnarray}
provided the drive and the response systems exists in identical piecewise linear regions. 
\end{subequations}
One can easily establish that a unique equilibrium point $ (x^{*}_0,y^{*}_0)$ exists for equation $(6)$ in each of the following three subsets,\\
\begin{equation}
\left.
\begin{aligned}
D^{*}_{+1} & =  \{ (x^{*},y^{*})| x^{*} > 1 \} P^{*}_+ = (0,0),\\
D^{*}_0 & =  \{ (x^{*},y^{*})|| x^{*} | < 1 \}| O^{*} = (0,0),\\
D^{*}_{-1} & =  \{ (x^{*},y^{*})|x^{*} < -1 \}| P^{*}_- = (0,0),\\
\end{aligned}
\right\}
\quad\text{}
\end{equation}
Naturally these fixed points can be observed depending upon the initial conditions $x^{*}_0$ and $y^{*}_0$ of Eqs.(5) which are inturn obtained from the initial conditions of the drive and response systems as, $x^{*}_0 = x_0 - x^{'}_0$ and $y^{*}_0 = y_0 - y^{'}_0$. In fact, Eqs.(5) can be integrated explicitly in terms of elementary functions in each of the three regions $D^{*}_0$, $D^{*}_{\pm}$ and the resulting solutions can be matched across the boundaries to obtain the full solution.\\The stability of the fixed points given in Eq.(7) can be calculated from the stability matrices. Hence, in the first case, $g(x)$ and $g(x^{'})$ takes the values  $a{x}$ and $a{x^{'}}$ respectively, corresponding to the central region in the $v-i$ characteristics of the nonlinear element, which is taken as the $D^{*}_0$ region of the difference system. In the $D^{*}_0$ region the stability determining eigen values are calculated from the stability matrix,
\begin{equation}
J^{*}_0 =
\begin{pmatrix}
-(a+\epsilon+1) &&& -1 \\
\beta &&& 0 \\
\end{pmatrix}
\end{equation}
The eigen values $m_1$ and $m_2$ are found to be a pair of complex conjugates for $\epsilon < 0.5682135$ while they are real and distinct for $\epsilon \ge 0.5682135$.\\
In the second case, $g(x)$ and $g(x^{'})$ takes the values  $bx \pm (a-b)$ and $bx^{'} \pm (a-b)$ respectively, corresponding to the outer regions in the $v-i$ characteristics of the nonlinear element, which is taken as the $D^{*}_{\pm1}$ regions of the difference system.
In the $D^{*}_{\pm1}$ regions, the stability determining eigen values are calculated from the stability matrix,
\begin{equation}
J^{*}_{\pm} =
\begin{pmatrix}
(b+\epsilon+1) &&& -1 \\
\beta &&& 0 \\
\end{pmatrix}
\end{equation}
The eigen values $m_1$ and $m_2$ are found to be a pair of complex conjugates for $\epsilon < 0.0519135$ while they are real and distinct for $\epsilon \ge 0.0519135$. The eigen values of the difference system in both the regions are thus determined by the value of the coupling parameter $\epsilon$. The eigenvalues of the difference systems in both regiona sre summarized in Table 1.
\begin{table}[H]
\begin{center}
\begin{tabular}{c|c|c|c|c}
\hline
Region		&  Fixed Point	&	 $\epsilon$				&		Eigenvalues   		& Stability						\\ \hline

$D^{*}_{0}$	&	(0,0)		&        $\epsilon < 0.5682135$ 		&		Complex conjugates	& Stable focus					\\
			&			&        $\epsilon \ge 0.5682135$		&		Real and distinct		& Stable node					\\ \hline
$D^{*}_{\pm1}$	&	(0,0)		&	$\epsilon < 0.0519135$		&		Complex conjugates	& Stable focus					\\
			&			&        $\epsilon \ge 0.0519135$		&		Real and distinct		& Stable node					\\ \hline
\end{tabular}
\caption{Stability of the fixed points in $D^{*}_{0}$ and $D^{*}_{\pm1}$ regions}
\label{tab1} 
\end{center} 
\end{table}
\begin{figure}[htb]
\begin{center}
\includegraphics[scale=0.66]{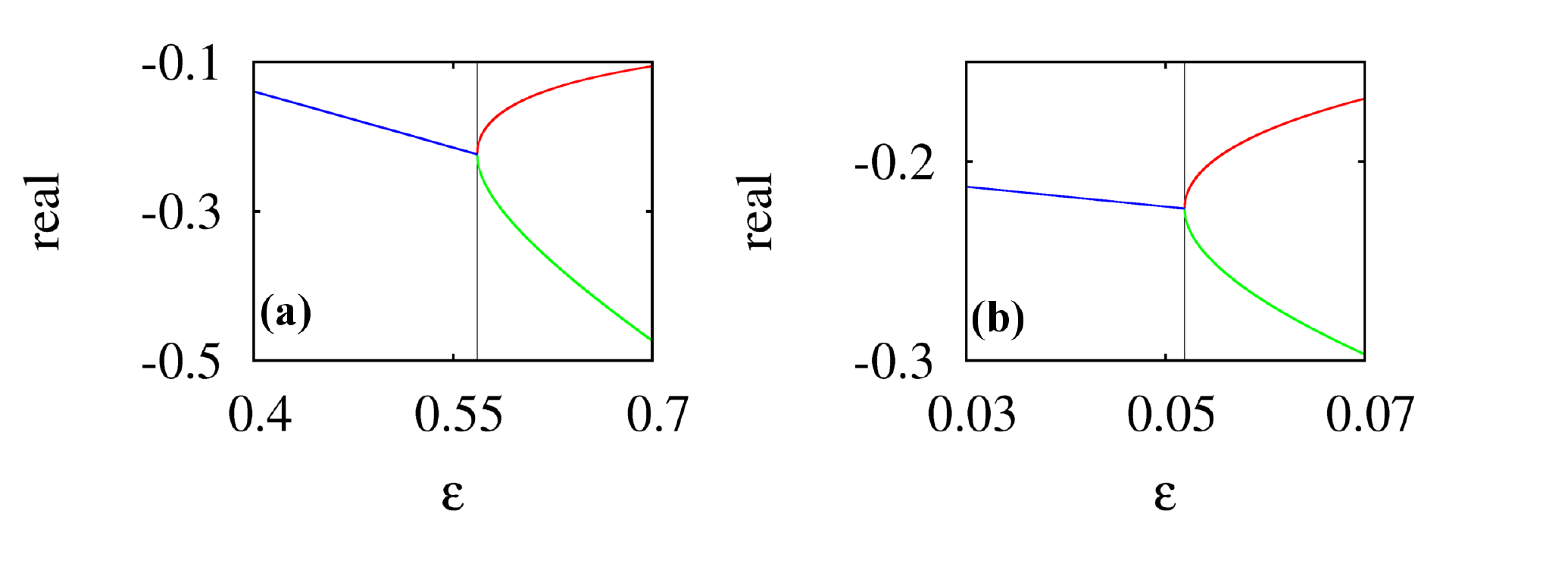}
\caption{Eigenvalues as functions of the coupling parameter $\epsilon$ in (a) $D^{*}_{0}$ region (b) $D^{*}_{\pm1}$ region}.
\label{fig:2a}
\end{center}
\end{figure}
Fig.4(a) $\&$ 4(b) shows the eigenvalues of the Jacobian matrix given by Eq.(8) and (9) in the $D^{*}_{0}$ and $D^{*}_{\pm1}$ regions respectively as functions of the coupling parameter $\epsilon$. The red and green lines show the two real roots while the blue line shows the real part of the complex conjugate roots as functions of the coupling parameter. In the next section, we present explicit analytical solutions for the dynamics of the response system driven by the drive system for value of the coupling parameter ($\epsilon > 0$) leading to complete synchronization of the drive and response.

\section{Explicit analytical solutions}

In this section, we present explicit analytical solutions for the unsynchronized $(\epsilon = 0)$ and synchronized states $(\epsilon > 0)$ of the coupled systems. When the coupling paramter $\epsilon=0$, the two systems are independent of each other. Hence the drive and the response systems given by equations (1) and (3) have the same solution for their state variables for all the three piecewise linear regions. 

\subsection*{Analytical solution for $\epsilon = 0$}

An explicit analytical solution for the single or uncoupled $MLCV$ circuit is presented and a chaotic attractor resembling that of the numerically obtained one as shown in Fig.(2) is observed in the $x-sin(\omega t)$ phase plane from the analytical solutions thus obtained \cite{Thamilmaran2001783}. The solutions of the normalized equations in terms of the state varaibles $y(t)$ and $x(t)$ obtained for each of the three piecewise linear regions $D_0$, $D_{\pm}$  are summarized as follows.
\subsection*{\ $Region: D_0$}
Since the roots $m_{1,2}$ in this region are a pair of complex congugates, the fixed point $(0,0)$ corresponding to the $D_0$ region is an {\emph{unstable spiral fixed point}}. The state variables $y(t)$ and $x(t)$ are
\begin{eqnarray}
y(t) &=& e^ {ut}(C_1 cos vt + C_2 sin vt) +E_1 sin(\omega_1 t)+ E_2 cos(\omega_1 t) \\
x(t) &=& \frac{1}{\beta}(\dot{y}) 
\end{eqnarray}

\subsection*{\ $Region: D_{\pm1}$}

The roots $m_{1,2}$ in this region are also a pair of complex conjugates. Hence the fixed points $k_1=0$, $k_2={\pm}(a-b)$ corresponding to the $D_{\pm1}$ region is a { \emph{stable spiral fixed point }}. The state variables $y(t)$ and $x(t)$ are
\begin{eqnarray}
y(t) &=& e^ {ut}(C_3 cos vt + C_4 sin vt) +E_3 sin(\omega_1 t)+ E_4 cos(\omega_1 t) {\pm} \Delta \\
x(t) &=& \frac{1}{\beta}(\dot{y}) 
\end{eqnarray}
where $\Delta = (b-a)$ and $+\Delta$, $-\Delta$ corresponds to $D_+$ and $D_-$ regions respectively.\\
The above solutions of the state variables can be used to obtain chaotic attractors for both the drive and response systems. Fig.5 shows the analytically obtained phase portraits for the chaotic attractors of the drive and response systems together in the $(x-y)$ and $(x^{'}-y^{'})$ phase planes for an amplitude $f_1=0.375, f_2 = 0.377$ and frequency $\omega_{1,2}=0.105$ of the external periodic force. Since the response system is operated with a different set of initial conditions and amplitude of the external periodic force, the chaotic attractors of the drive and response systems are non- identical and are unsynchronized. 
\begin{figure}[htb]
\begin{center}
\includegraphics[scale=0.5]{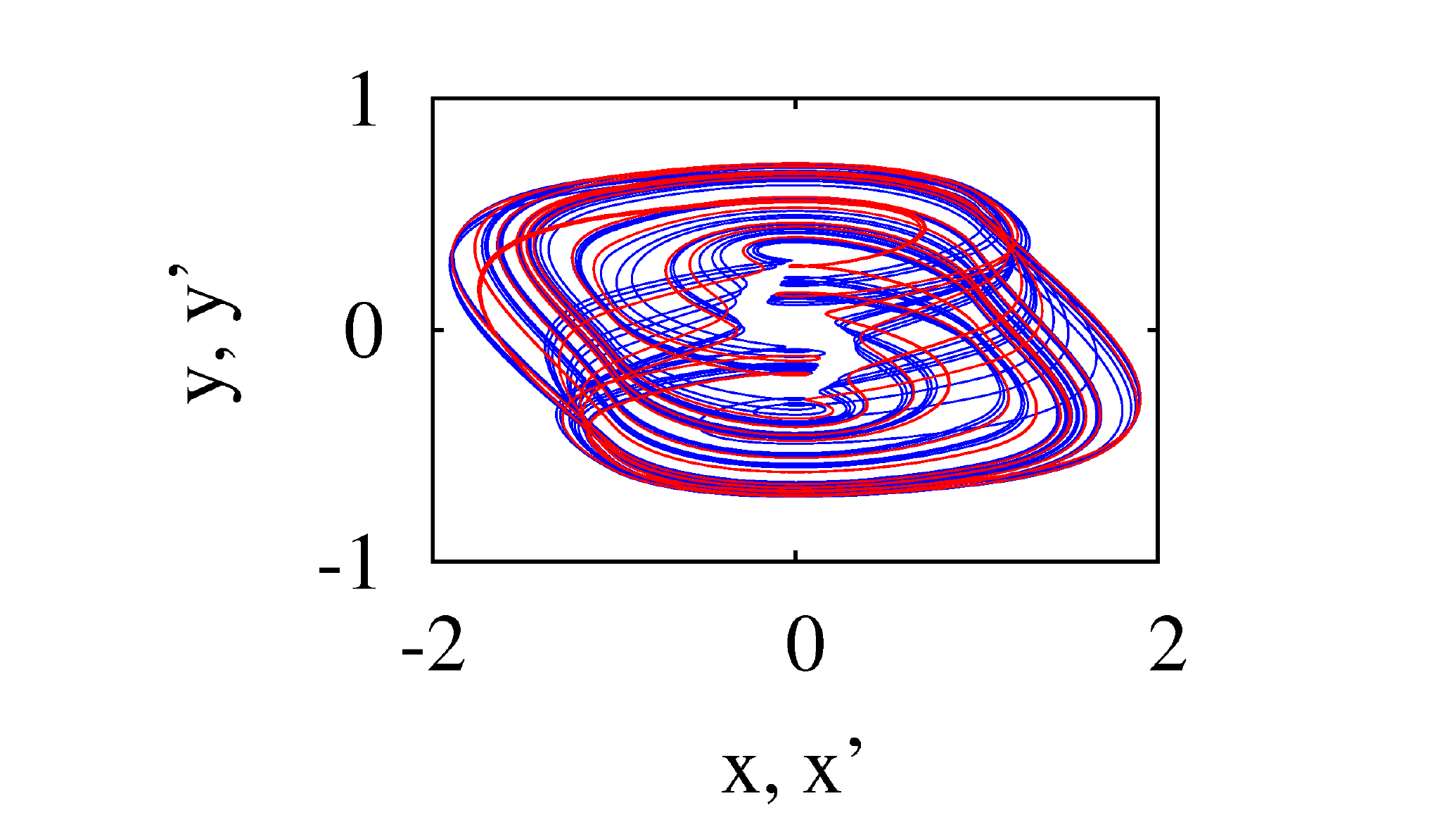}
\caption{Analytically obtained chaotic attractors of drive (blue line) and response (red line) systems showing the unsynchronized state of the two systems for $\epsilon = 0$.}
\label{fig:2a}
\end{center}
\end{figure}

\subsection*{Analytical solution for $\epsilon > 0$}

\vspace{-0.10cm} 

When the coupling parameter is increased ie.$(\epsilon >0)$, the response system is controlled by the drive and completely synchronizes with the drive for larger values of the coupling parameter. An explicit analytical solution to the dynamics of the response system could be obtained by finding a solution to the normalized state variables of the difference system given by Eq.5. The solution of those equations are, $ [x^{*} (t; t_0, x^{*}_0, y^{*}_0), ~y^{*}(t; t_0, x^{*}_0, y^{*}_0)]^T$ for which the initial conditions are written as $ (t, x^{*}, y^{*}) $ $ = (t_0, x^{*}_0, y^{*}_0) $. From the solution $x^{*}(t)$ and $y^{*}(t)$ thus obtained the state variables $x^{'}(t)$ and $y^{'}(t)$  can be found using Eqs.(6). Since Eq.(5) is piecewise linear, the solution in each of the three regions can be obtained explicitly. 

\subsection*{ \bf $Region: D^{*}_0$}

In this region $g(x)$ and $g(x^{'})$ takes the values  $a{x}$ and $a{x^{'}}$ respectively. Hence the normalized equations obtained from Eqs.(5) are,
\begin{subequations}
\begin{eqnarray}
\dot {x^{*}} & = & - (a+\epsilon+1)x^{*} - y^{*}+f_1 sin(\omega_1 t) - f_2 sin(\omega_2 t), \\
\dot {y^{*}} & = & \beta x^{*}
\end{eqnarray}
\end{subequations}
Differentiating Eq.~(14b) with respect to time and using Eqs.~(14a,~14b) in the resultant equation, we obtain
\begin{equation}
{\ddot y^{*}} + {A \dot y^{*}} + By^{*} = f_1 sin(\omega_1 t) - f_2 sin(\omega_2 t)
\end{equation}
where, $ A =  a + \epsilon +1$ and  $B = \beta$.
The roots of the equation (16) is given by
\begin{equation}
{m_{1,2}} =  \frac{-(A) \pm \sqrt{(A^{2}-4B)}} {2}, \nonumber
\end{equation}
\subsubsection*{\bf case: a}
For $\epsilon < 0.5682135$, $(A^{2} < 4B)$ and hence the roots $m_1$ and $m_2$ are a pair of complex congugates given as $m_{1,2} = u \pm iv$, with $u=\frac{-A}{2}$ and $v=\frac{\sqrt(4B-A^{2})}{2}$. The general solution to eq.(15) can be written as,
\begin{equation}
y^{*}(t) =  e^ {ut} (C_1 cosvt+ C_2 sinvt)+ E_1 sin \omega_1 t + E_2 cos \omega_1 t + E_3 sin \omega_2 t + E_4 cos \omega_2 t 
\end{equation}
where $C_1$ and $C_2$ are integration constants and
\begin{eqnarray}
E_1  &=&  \frac {f_1 \beta (B-{\omega_1}^2)}{A^2 {\omega_1} ^2 + (B-{\omega_1} ^2)^2} \nonumber \\
E_2  &=&  \frac {-A \omega_1 f_1 \beta}{A^2 {\omega_1} ^2 + (B-{\omega_1} ^2)^2} \nonumber \\
E_3  &=&   \frac {-f_2 \beta (B-{\omega_2}^2)}{A^2 {\omega_2} ^2 + (B-{\omega_2} ^2)^2} \nonumber \\
E_4  &=&  \frac {A \omega_2 f_2 \beta}{A^2 {\omega_2} ^2 + (B-{\omega_2} ^2)^2} \nonumber
\end{eqnarray}
Differentiating Eq.(16) and using it in Eq.(14b) we get,
\begin{equation}
x^{*}(t) = \frac{1}{\beta}(\dot{y^{*}})
\end{equation}
The constants $C_1$ and $C_2$ in the above equations can be evaluated by solving both Eqs.(16) and (17) for $C_1$ and $C_2$ at a suitable initial instant $t_0$, with $x^{*}_0$ and $y^{*}_0$ as initial conditions at time $t=t_0$, provided the trajectory of the dynamical system just enters the region $D^{*}_0$ at time $t_0$. The constants $C_1$ and $C_2$ thus obtained are \\
\begin{eqnarray}
C_1 =  \frac{e^ {- u t_0}} {v} \{(v { y^{*}_0} cos vt_0- (\beta { x^{*}_0}-u { x^{*}_0}) sin vt_0)+((\omega_1 E_1 - u E_2) sinvt_0 - vE_2 cosvt_0)cos \omega_1 t_0 \nonumber \\
             - ((\omega_1 E_2 +u E_1) sinvt_0+v E_1 cosvt_0) sin \omega_1 t_0 + ((\omega_2 E_3 - u E_4) sinvt_0 - vE_4 cosvt_0)cos \omega_2 t_0  \nonumber \\
	  - ((\omega_2 E_4 + u E_3) sinvt_0 + vE_3 cosvt_0) sin \omega_2 t_0 \}  \nonumber \\
C_2 =  \frac{e^ {- u t_0}} {v} \{((\beta { x^{*}_0}-u { x^{*}_0}) cos vt_0+v { y^{*}_0} sin vt_0)-((\omega_1 E_1 - u E_2) cos vt_0 + v E_2 sin vt_0)cos \omega_1 t_0 \nonumber \\
             + ((\omega_1 E_2 +u E_1) cos vt_0 - v E_1 sin vt_0) sin \omega_1 t_0 - ((\omega_2 E_3 - u E_4) cos vt_0 + vE_4 sin vt_0)cos \omega_2 t_0  \nonumber \\
	  + ((\omega_2 E_4 + u E_3) cos vt_0 - vE_3 sin vt_0) sin \omega_2 t_0 \} \nonumber 
\end{eqnarray}
From the results of $y^{*}(t),~x^{*}(t)$ obtained from eqs.(16), (17) and $y(t),~x(t)$ obtained from Eqs.(10), (11) $x^{'}(t)$ and $y^{'}(t)$ can be obtained from Eqs.(6).

\subsubsection*{ case: b}

For $\epsilon \ge 0.5682135$, $(A^{2} > 4B)$ and the roots $m_1$ and $m_2$ are real and distinct.
The general solution to eq.(16) can be written as,
\begin{equation}
y^{*}(t) = C_1 e^ {m_1 t} + C_2 e^ {m_2 t} + E_1 sin(\omega_1 t) + E_2 cos(\omega_1 t) + E_3 sin(\omega_2 t) + E_4 cos(\omega_2 t)
\end{equation}
where $C_1$ and $C_2$ are the integration constants and the constants $E_1,E_2,E_3,E_4$ are the same as in case(a).
Differentiating Eq.(18) and using it in Eq.(14b) we get,
\begin{equation}
x^{*}(t) = \frac{1}{\beta}(\dot{y^{*}})
\end{equation}
The constants $C_1$ and $C_2$ in the above equations can be evaluated by solving both Eqs.(18) and (19) for $C_1$ and $C_2$ at a suitable initial instant $t_0$, with $x^{*}_0$ and $y^{*}_0$ as initial conditions at time $t=t_0$, provided the trajectory of the dynamical system just enters the region $D^{*}_{0}$ at time $t_0$. The constants $C_1$ and $C_2$ are,
\begin{eqnarray}
C_1 =  \frac{e^ {- m_2 t_0}} {m_2 - m_1} \{ (\beta{ x^{*}_0} - m_1{ y^{*}_0}) + ( m_1 E_2 - \omega_1 E_1) cos \omega_1 t_0 + (\omega_1 E_2 + m_1 E_1) sin \omega_1 t_0 \nonumber \\
            +(m_2 E_4 - \omega_2 E_3) cos \omega_2 t_0 + (\omega_2 E_4 + m_2 E_3 ) sin \omega_2 t_0 \} \nonumber \\
C_2 =  \frac{e^ {- m_1 t_0}} {m_1 - m_2} \{ (\beta{ x^{*}_0} - m_2{ y^{*}_0}) + ( m_2 E_2 - \omega_1 E_1) cos \omega_1 t_0 + (\omega_1 E_2 + m_2 E_1) sin \omega_1 t_0 \nonumber \\
            +(m_1 E_4 - \omega_2 E_3) cos \omega_2 t_0 + (\omega_2 E_4 + m_1 E_3 ) sin \omega_2 t_0 \} \nonumber
\end{eqnarray} 
From the results of $y^{*}(t),~x^{*}(t)$ obtained from eqs.(18), (19) and $y(t),~x(t)$ obtained from Eqs.(10), (11) $x^{'}(t)$ and $y^{'}(t)$ can be odtained from Eqs.(6).

\subsection*{\bf $Region: D^{*}_{\pm1}$}

In this region $g(x)$ and $g(x^{'})$ takes the values  $b{x}\pm (a-b)$ and $b{x^{'}} \pm (a-b)$ respectively. Hence the normalized equations obtained from eqs. (5) are,
\begin{subequations}
\begin{eqnarray}
\dot {x^{*}} & = & - (b+\epsilon+1)x^{*} - y^{*}+f_1 sin(\omega_1 t) - f_2 sin(\omega_2 t), \\
\dot {y^{*}} & = & \beta x^{*}
\end{eqnarray}
\end{subequations}
Differentiating Eq.(20b) with respect to time and using Eqs.(20a,20b) in the resultant equation, we obtain
\begin{equation}
{\ddot y^{*}} + {C \dot y^{*}} + Dy^{*} = f_1 sin(\omega_1 t) - f_2 sin(\omega_2 t)
\end{equation}
where, $ C =  b + \epsilon +1$ and  $D = \beta$.
The roots of the equation (21) is given by
\begin{equation}
{m_{3,4}} =  \frac{-(C) \pm \sqrt{(C^{2}-4D)}} {2}, \nonumber
\end{equation}

\subsubsection*{case: a}

For $\epsilon < 0.0519135$, $(C^{2} < 4D)$ and hence the roots $m_3$ and $m_4$ are a pair of complex congugates given as $m_{3,4} = u \pm iv$, with $u=\frac{-C}{2}$ and $v=\frac{\sqrt(4D-C^{2})}{2}$. The general solution to eq.(16) can be written as,
\begin{equation}
y^{*}(t) =  e^ {ut} (C_3 cosvt+ C_4 sinvt)+ E_5 sin \omega_1 t + E_6 cos \omega_1 t + E_7 sin \omega_2 t + E_8 cos \omega_2 t 
\end{equation}
The constants $E_5,E_6, E_7, E_8$ are the same  as the contants $E_1, E_2, E_3, E_4$ in $case:a$ of $D^{*}_{0}$ region except that the constants $A  \& B$ are replaced with $C \& D$ respectively. 
Differentiating Eq.(22) and using it in Eq.(20b) we get,
\begin{equation}
x^{*}(t) = \frac{1}{\beta}(\dot{y^{*}})
\end{equation}
The constants $C_3$ and $C_4$ are the same as $C_1$ and $C_2$ in $case:a$ of $D^{*}_{0}$ region except that the constants $E_1, E_2, E_3, E_4$ are replaced with the constants $E_5,E_6, E_7, E_8$ respectively.
From the results of $y^{*}(t),~x^{*}(t)$ obtained from eqs.(22), (23) and $y(t),~x(t)$ obtained from Eqs. (12), (13) $x^{'}(t)$ and $y^{'}(t)$ can be obtained from Eq.(6).

\subsubsection*{\bf case: b}

For $\epsilon \ge 0.0519135$, $(C^{2} > 4D)$ and the roots $m_3$ and $m_4$ are real and distinct.
The general solution to eq.(24) can be written as,
\begin{equation}
y^{*}(t) = C_3 e^ {m_3 t} + C_4 e^ {m_4 t} + E_5 sin(\omega_1 t) + E_6 cos(\omega_1 t) + E_7 sin(\omega_2 t) + E_8 cos(\omega_2 t)
\end{equation}
where $C_3$ and $C_4$ are the integration constants and the constants $E_5,E_6,E_7,E_8$ are the same as in case(a).
Differentiating Eq.(24) and using it in Eq.(20b) we get,
\begin{equation}
x^{*}(t) = \frac{1}{\beta}(\dot{y^{*}})
\end{equation}
The constants $C_3$ and $C_4$ are the same as $C_1$ and $C_2$ in $case:a$ of $D^{*}_{0}$ region except that the constants $E_1, E_2, E_3, E_4$ are replaced with the constants $E_5,E_6, E_7, E_8$ respectively. From the results of $y^{*}(t),~x^{*}(t)$ obtained from eqs.(24), (25) and $y(t),~x(t)$ obtained from Eqs. (12), (13) $x^{'}(t)$ and $y^{'}(t)$ can be obtained from Eq.(6).\\

Now let us briefly explain how the solution can be generated in the $(x^{*}-y^{*})$ phase space. Thus if we start with the initial conditions $x^{*}(t=0) = x^{*}_0, y^{*}(t=0) = y^{*}_0$ in the region $D^{*}_0$ region at time $t=0$, the arbitrary constants $C_1$ and $C_2$ get fixed. Thus $x^{*}(t)$ evolves as given in Eq.(18) or Eq.(21), depending upon the value of $\epsilon$, up to either $t=T_1$, when $x^{*}(T_1)=1$ and $\dot{x^{*}}(T_1) > 0$ or $t=T^{'}_1$, when $x^{*}(T^{'}) = -1$ and $\dot{x^{*}}(T^{'}_1) < 0$. Knowing whether $T_1 < T^{'}_1$ or $T_1 > T^{'}_1$ we can determine the next region of interest $(D^{*}_{\pm1})$ and the arbitrary constants of the solutions of that region can be fixed by matching the solutions. The procedure can be continued for each successive crossing. In this way, the explicit solutions can be obtained in each of the regions $D^{*}_0$, $D^{*}_{\pm1}$. However, it is clear that sensitive dependence on initial conditions is introduced in each of these crossings at appropriate parameter regimes during the inverse procedure of finding $T_1, T^{'}_1, T_2, T^{'}_2,...,$ etc. from the solutions.
\begin{figure}[htb]
\begin{center}
\includegraphics[scale=0.6]{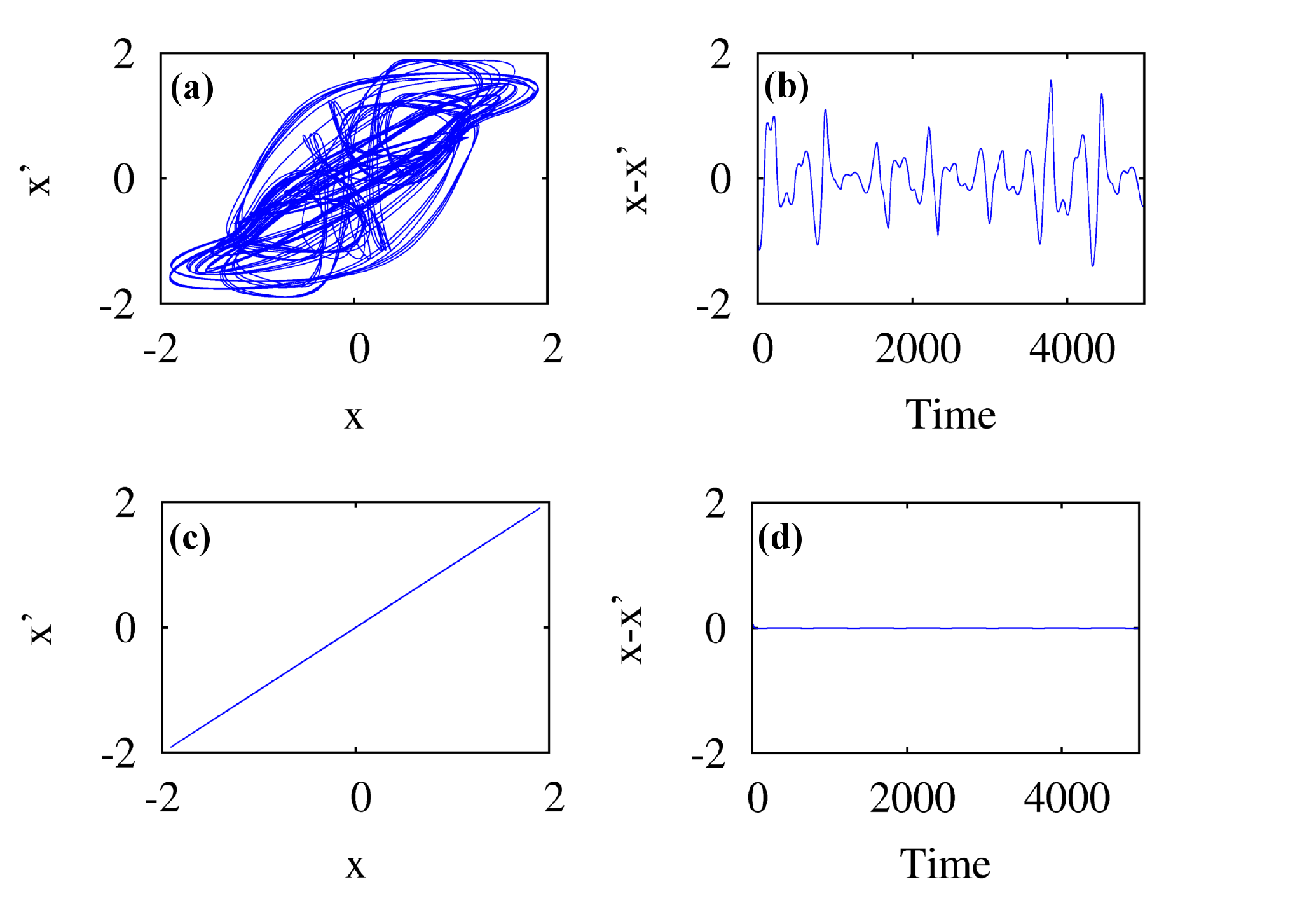}
\caption{(a) Unsynchronized motion in $(x-x^{'})$ plane for $\epsilon = 0.0$, $f_{1} = 0.375, f_{2} = 0.377$, [$x_0$=-0.5, $y_0$=0.1] and [$x^{'}_0$ = 0.5, $y^{'}_0$=0.11]. (b) Trajectory plot in $(x-x^{'})$ plane for $\epsilon=0$. (c) Synchronized motion in $(x-x^{'})$ plane for $\epsilon = 1.0$,  $f_{1} = 0.375, f_{2} = 0.377$, [$x_0$=-0.5, $y_0$=0.1] and [$x^{'}_0$ = 0.5, $y^{'}_0$=0.11]. (d) Trajectory plot in $(x-x^{'})$ plane for $\epsilon=1.0$}
\label{fig:2a}
\end{center}
\end{figure}
The analytical solutions obtained above for the response system can be used to explain the phenomena of complete synchronization through phase portraits. The amplitude and frequency of the drive and the response sytems are fixed as $f_{1}=0.375, f_2=0.377$ and $\omega_{1,2}=0.105$ respectively. Also the systems are operated with different set of initial conditions given by $x_0=-0.5, y_0=0.1$ and $x^{'}_0 = 0.5, y^{'}_0=0.11$ . Figure (6) shows the non-identical chaotic attractors of the drive and the response systems with the blue line giving the chaotic attarctor of the drive and the red line of the response. Fig.6 shows the Unsynchronized and Synchronized states of the coupled systems for two different values of the coupling parameter $\epsilon$. Figure 6(a) shows the unsynchronized state of the coupled systems for the coupling parameter $\epsilon=0$ in the $(x-x{'})$ phase plane and the corresponding trajectory in the $x^{*}=x-x^{'}$ plane. As the value of the coupling parameter is increased, the response system completely synchronizes with the drive. Figure 6(c) shows the complete synchronization of the coupled systems in the $(x-x{'})$ phase plane for the value of the coupling parameter $\epsilon = 1.0$ and the corresponding trajectory in the $x^{*}=x-x^{'}$ plane as in Fig.6(d). From the phase portraits and the time series plots obtained it could be inferred that for the coupling parameter taking the value $\epsilon=1.0$, the response system which is operating with a different set of initial condition and for a different value of external periodic force, completely synchronizes with the drive. 

\section{Conclusion}
In this paper, we presented numerical and explicit analytical studies on unidirectionally coupled two variant of $MLC$ circuits exhibiting complete synchronization in their dynamics. The complete synchronization of the response system which possesses a non-identical choatic attractor with that of the drive, for a particular value of the coupling parameter is studied through phase portraits obtained numerically and analytically. 

\section*{Acknowledgement}
This work is supported by {\emph{Alagappa Chettiar College of Engineering $\&$ Technology}} under {\emph{Technical Education Quality Improvement Programme(TEQIP)-II}} scheme.

\bibliographystyle{plain}
\bibliography{mybib}

\end{document}